\documentclass[sigconf]{acmart}
\AtBeginDocument{%
  \providecommand\BibTeX{{%
    \normalfont B\kern-0.5em{\scshape i\kern-0.25em b}\kern-0.8em\TeX}}}

\setcopyright{acmcopyright}
\copyrightyear{2023}
\acmYear{2023}
\acmDOI{XXXXXXX.XXXXXXX}

\acmConference[ICONS 2023]{International Conference on Neuromorphic Systems 2023}{Aug 1--3, 2023}{Santa Fe, NM, USA}

\acmPrice{15.00}
\acmISBN{978-1-4503-XXXX-X/18/06}

\usepackage{booktabs}
\usepackage{multirow}
\newcommand{\family}{$\mathcal{F}$}

\begin{document}

\title{Memristive Reservoirs Learn to Learn}

\author{Ruomin Zhu}
\affiliation{%
  \institution{School of Physics, \\The University of Sydney}
  \city{Sydney}
  \state{NSW}
  \country{Australia}
  \postcode{2006}
}
\email{rzhu0837@uni.sydney.edu.au}

\author{Jason K. Eshraghian}
\affiliation{%
  \institution{Department of Electrical and Computer Engineering, \\University of California, Santa Cruz}
  \city{Santa Cruz}
  \state{CA}
  \country{USA}
  \postcode{95064}
}
\email{jeshragh@ucsc.edu}

\author{Zdenka Kuncic}
\affiliation{%
  \institution{School of Physics and 
  \\Sydney Nano Institute,
  \\The University of Sydney}
  \city{Sydney}
  \state{NSW}
  \country{Australia}
  \postcode{2006}
}
\email{zdenka.kuncic@sydney.edu.au}



\begin{abstract}
Memristive reservoirs draw inspiration from a novel class of neuromorphic hardware known as nanowire networks. These systems display emergent brain-like dynamics, with optimal performance demonstrated at  dynamical phase transitions. 
In these networks, a limited number of electrodes are available to modulate system dynamics, in contrast to the global controllability offered by neuromorphic hardware through random access memories.
We demonstrate that the learn-to-learn framework can effectively address this challenge in the context of optimization. 
Using the framework, we successfully identify the optimal hyperparameters for the reservoir. This finding aligns with previous research, which suggests that the optimal performance of a memristive reservoir occurs at the `edge of formation' of a conductive pathway.
Furthermore, our results show that these systems can mimic membrane potential behavior observed in spiking neurons, and may serve as an interface between spike-based and continuous processes.
\end{abstract}




\received{20 February 2007}
\received[revised]{12 March 2009}
\received[accepted]{5 June 2009}

\maketitle

\section{Introduction}

Nanowire networks are a novel class of neuromorphic devices that demonstrate potential for brain-inspired computing and information processing \cite{Stieg2012, Sillin2013, Demis2015, Kuncic2020, Zhu2020harnessing, Lilak2021, Loeffler2021, Zhu2021information, Hochstetter2021, Zhu2021mnist, Kuncic2021, Milano2021, Loeffler2023}.
By embedding memristive switching dynamics into naturally-arising neuromorphic connectivity structures~\cite{Loeffler2020, Milano2022}, these networks display emergent brain-like  dynamics \cite{Diaz-Alvarez2019}, including dynamical phase transitions and avalanche criticality \cite{Hochstetter2021, Dunham2021}.

The synaptic sites of nanowire networks are not directly accessible, in contrast to random access memories (RAM) \cite{Chang2016, eshraghian2022memristor, ielmini2018memory}, where each memory cell is addressable and programmable.
The lack of controllability is compensated for by the dynamic nature of nanowire networks, which is a key feature that enables them to adapt to evolving input signals.
Nevertheless, it is worth investigating how these neuromorphic systems can be optimized for information processing tasks. For example, previous studies have shown that in a physical reservoir computing framework, nanowire networks can achieve superior learning performance when operating near a dynamical phase transition~\cite{Hochstetter2021, Zhu2021information}. 
Rather than manually exploring the optimal region of operation, an alternative and more effective way to optimize the parameters that can be physically adjusted (namely, inputs and outputs) is the learn-to-learn (L2L) framework, commonly known as meta-learning.

The L2L approach is a scheme for optimizing learning capacity from prior experiences \cite{Hospedales2020}. 
Recent advances of L2L algorithms are built upon the premise that the learning system -- typically an artificial neural network -- is differentiable, so that the gradients can be propagated through the network to adjust the hyperparameters (e.g. learning rate) \cite{Hochreiter2001, Andrychowicz2016, Finn2017}. 
It has also been shown that biologically or physically inspired systems are suitable candidates as learning agents in the L2L framework \cite{Bellec2018, Subramoney2021}.
In particular, Bohnstingl et al. \cite{Bohnstingl2019} showed that non-differentiable systems could be optimized using non-gradient-based optimization schemes.

This study demonstrates how learning is achieved using the collective dynamics of a system abstracted from physical nanowire networks (see details in~\cite{Hochstetter2021, Zhu2021information}) under a physical reservoir computing (RC) framework~\cite{JaegerESN, Maass2002}, namely a memristive reservoir, where training is restricted to the readout layer to circumvent the computation burden of training traditional deep artificial network architectures~\cite{Lukosevicius2009}. The synaptic sites (recurrent weights) in the reservoir are not individually programmable but this may be offset by the highly rich set of dynamics available in the memristive substrate as a response to external stimuli. 
The L2L approach is applied to the memristive reservoir and we show that this system is able to learn the dynamics of a family of nonlinearly filtered signals.
Furthermore, we also demonstrate that the memristive reservoir system is able to generate dynamics that resemble the membrane potentials of spiking neural networks (SNNs) directly from continuous inputs, which implies its potential to bridge the gap between continuous signals and spike-based computing paradigms.
By learning across various membrane potential dynamics, meta-learning can identify common principles or shared features that govern the dynamics of both spiking neurons and nanowire networks, which may offer deeper insights into the underlying ionic mechanisms that pervade both biology and emerging memory technologies.
In particular, a nested-loop structure is implemented as following:
\begin{itemize}
    \item L2L is used in the outer loop to determine the optimal hyperparameters of the memristive reservoir;
    \item Linear regression is employed in the inner loop to optimize the stimulus/readout protocol to achieve the task objectives.
\end{itemize}

Demonstrating how L2L is compatible with systems grounded in physics, such as memristive reservoirs, opens up the potential to optimize learning using adaptive networks where individual internal weights are not trained, but rather allowed to self-adjust in repsonse to dynamical inputs, in a manner similar to the brain's neural network.
This has implications beyond memristive reservoirs and could lead to more efficient and effective optimization schemes for learning complex physical phenomena.





\section{Research Methods}
\begin{figure}[h]
  \includegraphics[width=\linewidth]{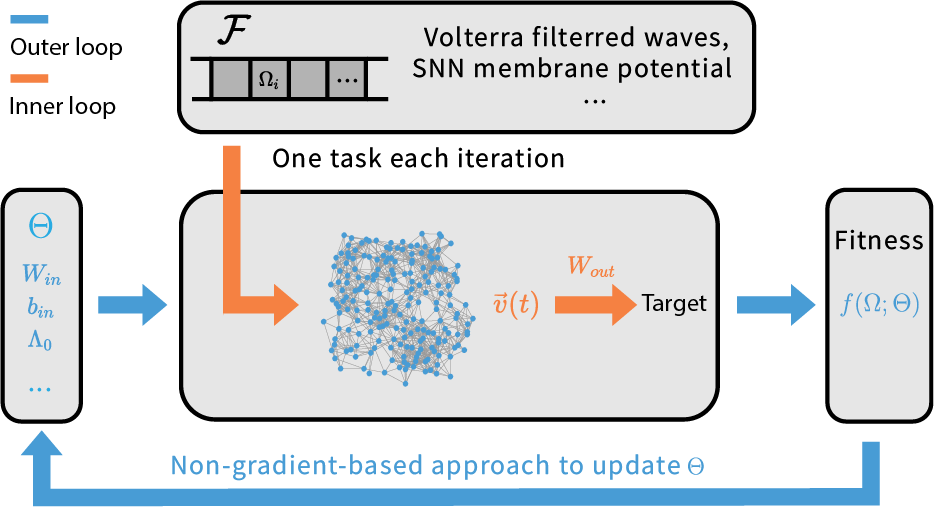}
  \caption{Schematic diagram of the L2L framework. 
    At each iteration of the outer loop, the hyperparameter set $\Theta = \{W_{in}, b_{in}, \Lambda_0, ...\}$ is perturbed in multiple directions.
    These perturbed hyperparameters are passed to the inner loop and learning task $\Omega$ randomly selected from \family are performed. 
    $\Theta$ is then optimized in the outer loop based on the returned fitness of different tasks. 
    The inner loop operations are indicated by orange arrows and the outer loop operations by blue arrows.
  }
  \label{fig:schematic}
\end{figure}

As continuous temporal input voltage signals are delivered to a memristive reservoir via dedicated input nodes, the network autonomously adjusts its internal state. 
Voltages of a subset of the remaining nodes can be read out and used to train weights in the fully-connected linear output layer.
Two L2L task families are studied: learning nonlinear Volterra dynamics and learning membrane dynamics from SNNs.

\subsection{The L2L Framework}
Fig.~\ref{fig:schematic} illustrates the L2L framework, where two iterative loops work together to optimize the system's learning performance for a chosen family of tasks \family.
In this work, the task families are split into two subsets, one used for meta-training and the other for meta-testing. 

The objective of the outer loop (orange in Fig.~\ref{fig:schematic}) is to optimize the reservoir hyperparameters $\Theta$ based on the fitness in the inner loop. 
In each iteration of the inner loop (blue in Fig.~\ref{fig:schematic}), the system learns from a specific task $\Omega_i$ within the meta-training phase of \family and the fitness $f(\Omega_i;\Theta)$ is evaluated.
Non-gradient-based techniques, namely simulated annealing (SA) and evolutionary strategies (ES), are applied to the fitness to determine the optimal hyperparameter set $\Theta'$ (or range), thus achieving the best collective learning outcome for \family:
\begin{align}
    \Theta' = \underset{\Theta}{\text{argmax}} (f(\Omega;\Theta)), \quad \Omega \in \mathcal{F}.
\end{align}
After 100 generations of outer-loop training, the performance of the baseline (without meta-learning) and meta-learned reservoirs are evaluated using tasks from the meta-testing set to evaluate the influence of the L2L algorithm on the reservoir's learning performance.

\subsubsection{Simulated Annealing}
Simulated annealing  optimizes the objective function by mimicking the physical annealing process \cite{Kirkpatrick1983}. The algorithm is parameterized by a decaying `temperature' $T$.
At each generation of the outer loop, the hyperparameters are perturbed in multiple directions by a step size $\epsilon$ drawn from a normal distribution parameterized by $T$. 
For each perturbed hyperparameter $\tilde{\Theta}$, the change in fitness ($\Delta f = f(\Omega, \tilde{\Theta}) - f(\Omega;\Theta)$) is estimated.  $\tilde{\Theta}$ is accepted if the corresponding fitness improves, otherwise the probability ($P$) of accepting $\tilde{\Theta}$ is determined by an exponential distribution:
\begin{align}
    P = \begin{cases}
    1,                   &\Delta f \geq 0 ,\\
    e^{\Delta f / T},    &\Delta f < 0.
    \end{cases}
\end{align}
When $T$ is high, some sub-optimal solutions can be accepted and the algorithm explores a broader parameter space. As it `cools down', the probability of accepting worse solutions decreases and the solution eventually converges if a global optimum exists.

\subsubsection{Evolution Strategies}
Evolution strategies are inspired by natural selection and evolution \cite{Rechenberg1973, Wierstra2008}. A population of candidate solutions are generated and evolved based on weighted fitness and a pre-determined learning rate $\eta$ \cite{Sehnke2010, Salimans2017}. 
At generation $k$, each candidate $\Theta_{k,i}$ in the population is perturbed by a step size $\epsilon_i$ drawn from a normal distribution $(0,\, \mathbb{I}\,)$, with standard deviation $\sigma$ 

After evaluating the fitness $f(\Omega_i; \Theta_{k,i}+\sigma\epsilon_i)$ for all candidates in the same generation, $\Theta_{k,i}$ is evolved by:
\begin{align}
    \Theta_{k+1, i} = \Theta_{k,i} + \frac{\eta}{n \sigma} 
                        \sum_{i=1}^n f(\Omega_i; \Theta_{k,i}+\sigma\epsilon_i)\epsilon_i,
\end{align}
where $n$ is the size of the population.
The statistical setup enables ES to deal with high-dimensional problems, making it useful for optimization problems with many parameters. 

\subsection{Learning Volterra dynamics}
\begin{figure*}[h]
  \includegraphics[width=0.9\textwidth]{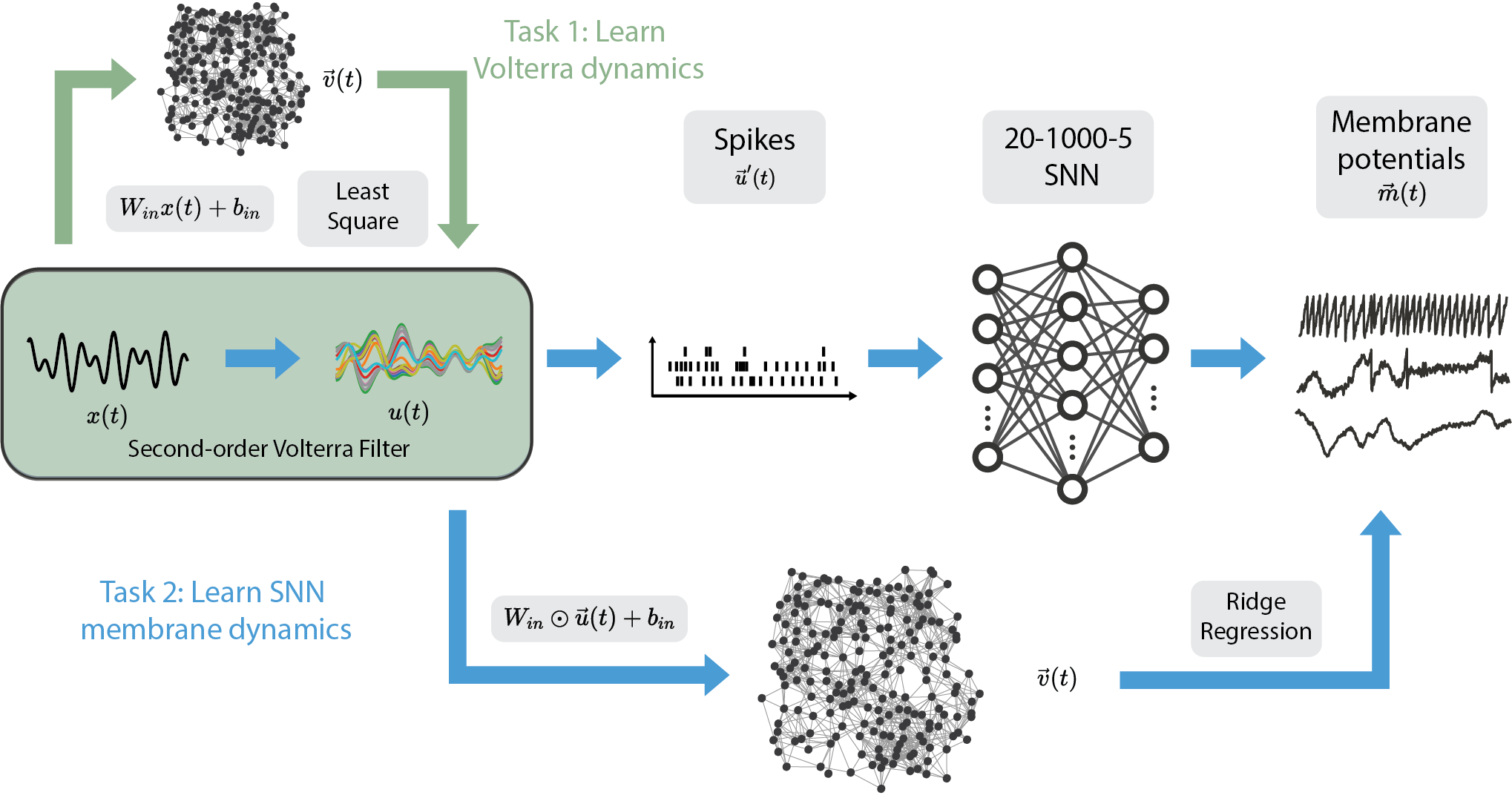}
  \caption{
  Schematic diagram for the learning tasks employed in this study. 
  The green box and arrows summarize the learning Volterra dynamics task family (Task 1), where the original signal $x(t)$ is the input and the Volterra-filtered signal $u(t)$ is the target for the reservoir.
  The blue arrows illustrate the flow of the learning SNN dynamics task family (Task 2), in which a set of 20 Volterra-filtered signals $\vec{u} (t)$ is converted to spikes $\vec{u}'(t)$ and delivered to a fully connected SNN. The membrane potentials $\vec{m}(t)$ of the readout neurons are used as target while the set of continuous Volterra-filtered signals are delivered directly to the reservoir as input.
  }
  \label{fig:task_schematic}
\end{figure*}

Consider a two-terminal configuration of a memristive reservoir with one source node and one drain node (see \cite{Zhu2021information} for details).
Fig.~\ref{fig:task_schematic} shows how the memristive reservoir is used to learn the nonlinear Volterra dynamics (Task 1).
As the green box and arrows indicate, the input signal $x(t)$ is delivered to the source node, while the drain node is grounded and an external fully connected output layer linearly combines the voltage read outs $\vec{v}(t)$ of 64 other nodes to regress to each target signal $u(t)$. 
The objective of the L2L process here is to fine-tune the hyperparameters to achieve optimal regression for the nonlinear time-delayed target signal.
The input weight ($W_{in}$), input bias ($b_{in}$), and initial reservoir state ($\Lambda_0$) are optimized by L2L via simulated annealing. 
$\Lambda_0$ is parameterized by a pulse of 1\,V DC with varying width $T_0$ applied to the input node prior to the task, as described in \cite{Zhu2021information}. 

The input signal $x(t)$ is generated by combining two sine waves:
\begin{align}
    x(t) = A_1 \sin (2 \pi \frac{t}{T_1} + \phi_1) + A_2 \sin (2 \pi \frac{t}{T_2} + \phi_2),
\end{align}
where $T_1 = 0.323$\,s and $T_2 = 0.5$\,s, and where $A_1, A_2 \in [0.5,1]$ and $\phi_1, \phi_2 \in [0,\frac{\pi}{2}]$ are chosen randomly.
The target signal $u(t)$ is generated using a second-order Volterra filter:
\begin{align}
    u(t)=&\int_\tau k_\Omega^1(\tau) x(t-\tau) d \tau \quad +  \nonumber \\ 
        &\int_{\tau_1} \int_{\tau_2} k_\Omega^2\left(\tau_1, \tau_2\right) 
        x \left(t-\tau_1\right) x\left(t-\tau_2\right) d \tau_1 d \tau_2,
\end{align}
in which $\tau, \tau_1, \tau_2 \in [0,0.5]$ denote the time delay of the signals, while $k_\Omega^1$ and $k_\Omega^2$ are task-specific random Volterra kernels (see \cite{Subramoney2021} for further information on generating the kernels). 
100 target signals are generated using different Volterra kernels to comprise the family \family.
The green box in Fig.~\ref{fig:task_schematic} shows an example of one input signal (black) and Volterra-filtered target signals (colored) with different random kernels.

The readout weights $W_{out}$ are trained using least square to minimize the loss:
\begin{align}
    \mathcal{L} = || U - W^\intercal V||_2,
\end{align}
where $||\cdot||_2$ represents the L-2 norm, and $U = [u(1), u(2), ... u(t)]$ and $V = [\vec{v}(1), \vec{v}(2), ..., \vec{v}(t)]$ are the stacked target and readout of the reservoir.
The learning results are estimated by normalized root mean squared error (NRMSE):
\begin{align}
    \text{NRMSE} = \frac{\sqrt{\frac{1}{T} \sum_{t=1}^{T}{(W_{out}^\intercal \vec{v}(t) - u(t))^2}}}{u_{max} - u_{min}},
\end{align}
where $T$ is the length of the signal.

\subsection{Learning membrane dynamics of SNNs}

The blue arrows in Fig.~\ref{fig:task_schematic} illustrate the scheme for this task family. A set of 20 randomly generated Volterra filter signals $\vec{u}(t)$ are utilized as input and converted to spike trains $\vec{u}'(t)$ using delta modulation. 
A fully connected SNN ($\mathcal{S}$, developed using~\cite{Eshraghian2021}) with 20 input neurons, 1,000 hidden neurons, and 5 output neurons receives $\vec{u}'(t)$ as input. 
For each task, the internal weights $W_{\mathcal{S}}$ of the SNN are randomly generated, and the membrane potentials of the 5 output neurons, $\vec{m}(t)$ are employed as the target signals of the learning task:
\begin{align}
    \vec{m}(t) = \mathcal{S} (\vec{u}'(t);W_{\mathcal{S}}).
\end{align}

The continuous Volterra signals $\vec{u}(t)$ are delivered to 20 input nodes in the memristive reservoir ($\mathcal{R}$), and 64 nodes are used as voltage readouts:
\begin{align}
    \vec{v}(t) = \mathcal{R}(W_{in} \odot \vec{u}(t) + \vec{b}_{in} ; \Lambda_0).
\end{align}
Notice that $W_{in}$ and $\vec{u}(t)$ have the same dimensions and $\odot$ represents element-wise multiplication.

For each task, the entire data stream is divided into three parts: 
the first 2000 data points are considered as a transient phase~\cite{Jaeger2002};
the subsequent 6000 data points (from 2000th to 8000th) employed as the support set (inner-loop training), 
and the last 1000 data points (from 8000th to 9000th) comprise the query set (inner-loop testing).
The readout weights $W_{out}$ are trained for each task using ridge regression to minimize the loss function:
\begin{align}
    \mathcal{L} = ||M - W_{out}^\intercal V||^2_2 + \alpha ||W_{out}||^2_2, 
\end{align}
where $\alpha$ is a regularization term ranging between $10^{-4}$ and 1, with 
$M = [\vec{m}(1), \vec{m}(2), ..., \vec{m}(t)]$ representing the stacked target membrane potentials.
A 5-fold cross validation scheme is employed to avoid overfitting \cite{hastie2009elements}. NRMSE of the query regime is reported as result.


A family \family of 150 tasks is created by utilizing SNNs with different internal weights.
The goal of L2L in this context is to adapt the reservoir to new tasks by extrapolating information from past learning experiences, allowing it to generate desired membrane potentials.
Simulated annealing and evolution strategies are used separately to optimize the input weight ($W_{in}$) and the input bias ($\vec{b}_{in}$) of each signal.
$\Lambda_0$ is fixed to the phase transition regime found in task 1.


\section{Results}
\subsection{L2L Volterra dynamics}
\begin{figure}[h]
    \centering
    \includegraphics[width=\linewidth]{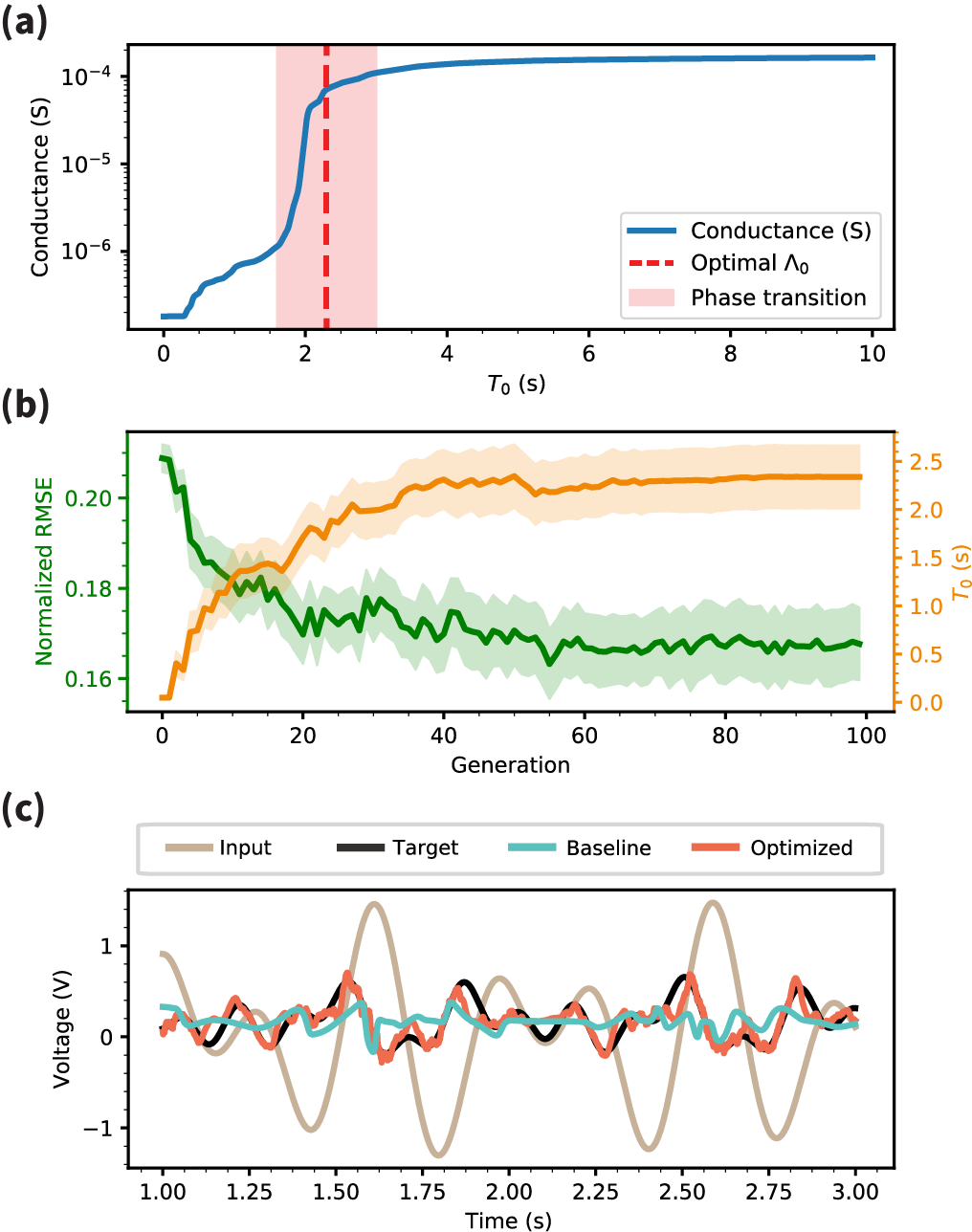}
    \caption{L2L Volterra dynamics.
            (a) Memristive reservoir conductance response (blue) to a varying 1\,V DC pulse width $T_0$, showing a characteristic phase transition regime (shaded). The dashed red line represents the reservoir's optimal initial state for learning Volterra dynamics, as found by the L2L scheme. 
            (b) Normalized RMSE of the learning tasks and optimal $T_0$ with respect to outer loop generation number.
            (c) Input signal, corresponding target, and the resulting learning outcomes for one representative task in the family (baseline is before optimization).
            }
    \label{fig:volterra}
\end{figure}

\begin{table}[h]
\caption{NRMSE for Task 1 with and without meta-learning}
\begin{tabular}{c|cc}
                  & Meta-train      & Meta-test       \\ \hline
w/o meta-learning & 0.209$\pm$0.003 & 0.211$\pm$0.014 \\
Meta-learned (SA) & 0.168$\pm$0.008 & 0.164$\pm$0.009 \\
\end{tabular}
\label{table:volterra}
\end{table}

Fig.~\ref{fig:volterra}(a) shows the reservoir's conductance (blue curve) in response to a 1\,V DC pulse of varying width $T_0$.
The shaded region represents the general phase transition regime, identified from previous studies \cite{Zhu2021information}, and the dashed line at $T_0=2.3$\,s indicates the optimal initial reservoir state $\Lambda_0$ found by the SA optimization scheme, as shown in Fig.~\ref{fig:volterra}(b).
$T_0$ (orange) converges toward $\simeq 2.3\,$s as RNMSE (green) converges to a minimum after approximately 60 generations.
Fig.~\ref{fig:volterra}(c) compares the system's learning outcomes for a specific task in the Volterra family before (blue) and after (red) the outer loop optimization (cf. Fig.~\ref{fig:schematic}) and Table.~\ref{table:volterra} shows the corresponding NRMSE of the whole meta-testing set.
A notable improvement in regression to the target signal is evident.

These results demonstrate that the L2L framework is successful in fine-tuning the reservoir's initial state $\Lambda_0$.
Remarkably, the optimization scheme finds the optimal $\Lambda_0$ coinciding with the dynamical phase transition region identified independently in previous studies \cite{Zhu2021information} as producing optimal task performance from memristive reservoirs.

\begin{figure}[h]
    \centering
    \includegraphics[width=\linewidth]{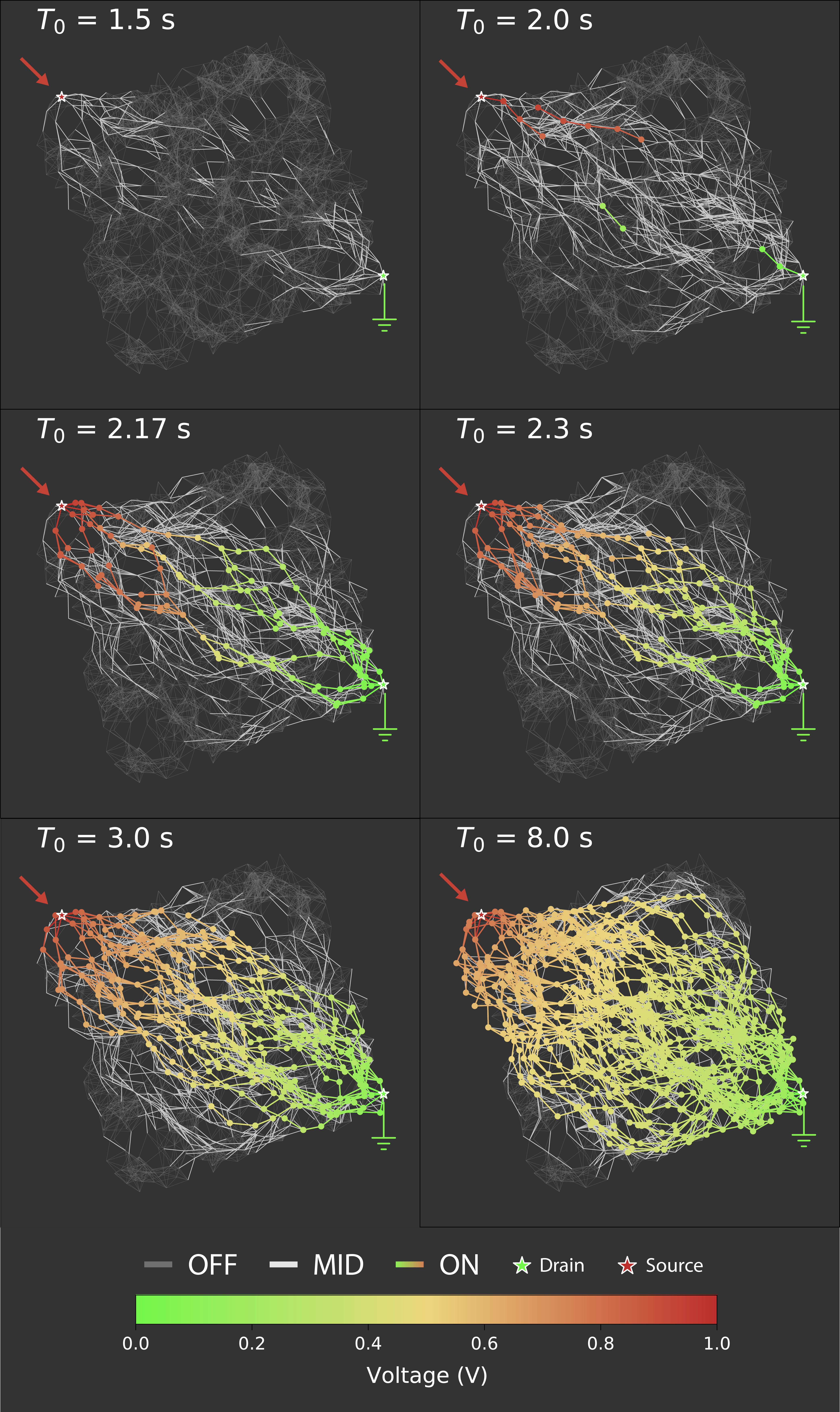}
    \caption{Reservoir activity with different pre-initialization pulse widths. 
    A 1\,V DC signal of varying width is delivered to the reservoir in prior to the task.
    Nodes on the conductance path from source to drain are colored based on their voltages.
    Edges are catagorized as high (ON), intermediate (MID) and low (OFF) conductance levels and colored respectively.
    }
    \label{fig:snapshots}
\end{figure}

\begin{figure}[h]
    \centering
    \includegraphics[width=0.8\linewidth]{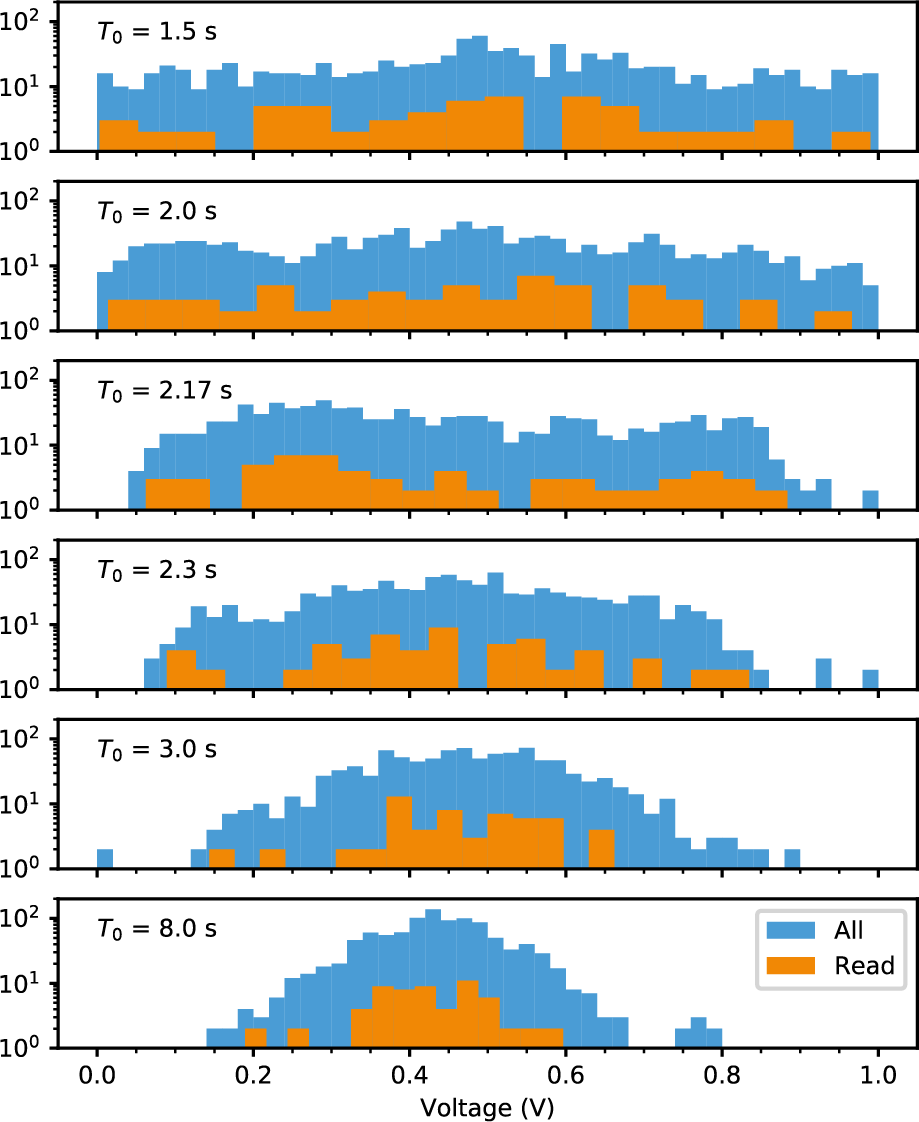}
    \caption{Reservoir node voltage histograms for different pre-initialization pulse widths $T_0$. Blue indicates all nodes, orange indicates nodes used as readouts for the learning tasks.}
    \label{fig:histos}
\end{figure}

To gain deeper insight into the internal dynamics of memristive reservoirs, Fig.~\ref{fig:snapshots} shows visualisation snapshots of network graphs of the reservoir for different $T_0$. 
When the reservoir is under-activated ($T_0 < 2\,$s), most memristive components are inactive and not enough information can be extracted to perform learning. 
On the other hand, when the reservoir is over-activated ($T_0 > 8$\,s), the internal dynamics saturates.%
The initial reservoir state at $T_0 \approx 2.17$\,s results in the best task performance and qualitatively, it is evident from Fig.~\ref{fig:snapshots} that this corresponds to an intermediate state, where conductance paths first span the network.
At this `edge of formation', the internal state of the reservoir produces voltage readout features that are more diverse than at later activation times, as shown in the corresponding node voltage distributions in Fig.~\ref{fig:histos}.

\subsection{L2L SNN dynamics}

\begin{table}[h]
\caption{NRMSE for Task 2 with and without meta-learning}
\begin{tabular}{c|cc}
                  & Meta-train      & Meta-test       \\ \hline
w/o meta-learning & 0.212$\pm$0.041 & 0.235$\pm$0.036 \\
Meta-learned (SA) & 0.139$\pm$0.011 & 0.164$\pm$0.012 \\
\textbf{Meta-learned (ES)} & \textbf{0.109$\pm$0.011} & \textbf{0.128$\pm$0.004}
\end{tabular}
\label{table:snn}
\end{table}

\begin{figure}[h]
    \centering
    \includegraphics[width=\linewidth]{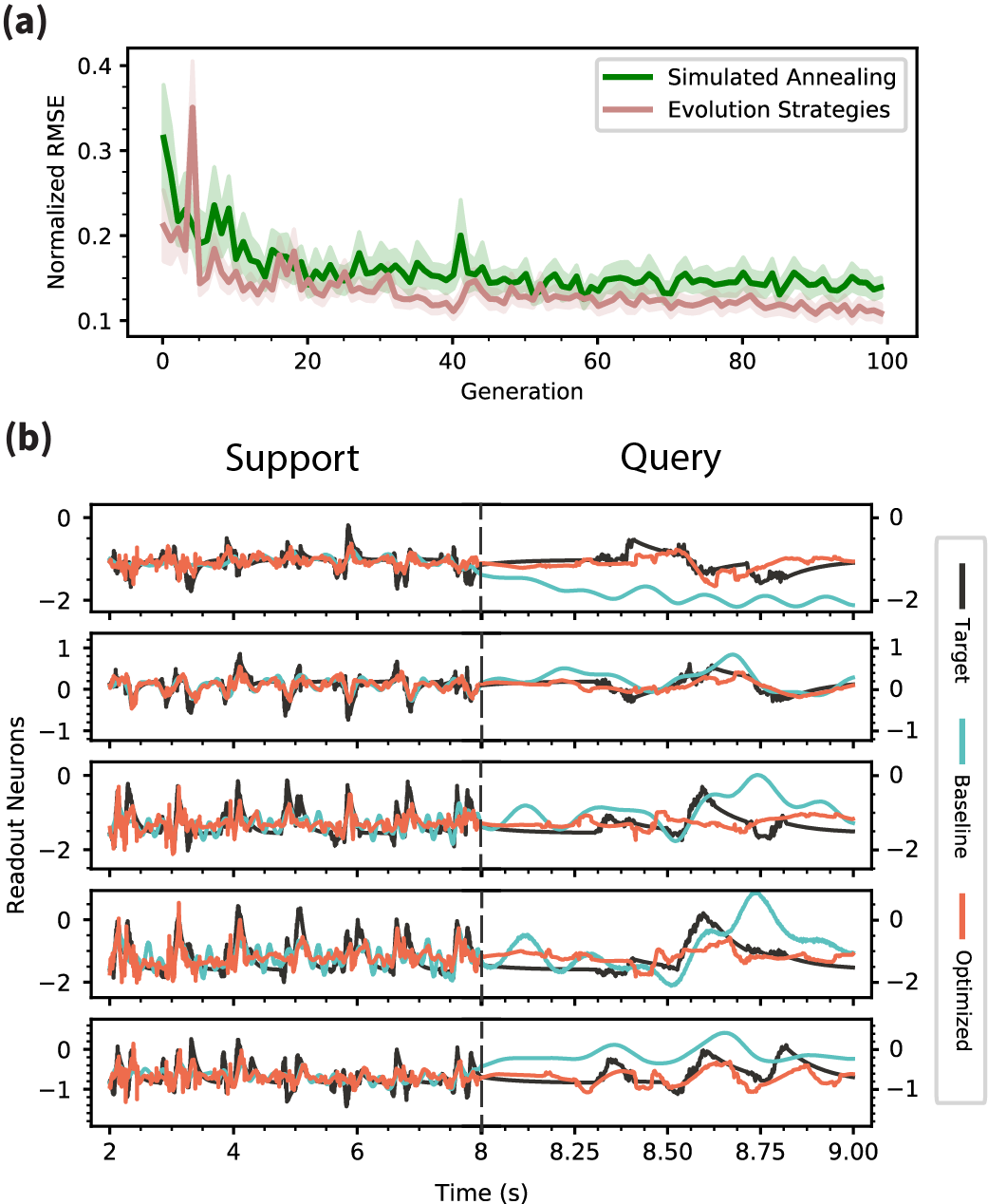}
    \caption{L2L SNN membrane dynamics. 
            (a) Normalized RMSE of the learning tasks as a function of generations in the outer loop.
            (b) Membrane potentials of 5 readout neurons from the SNN (black) overlaid by the learned dynamics from the baseline (blue) and the optimized (orange) memristive reservoirs for one representative meta-testing task in the task family.
            The left and right columns of panels show results for the support and query periods of the task, respectively. Note: a different timescale is used during the query phase for better visibility at test time.
            }
    \label{fig:learn_snn}
\end{figure}

Fig.~\ref{fig:learn_snn}(a) shows the NRMSE of the query part (task-specific inner loop testing) for Task 2 during outer loop training for the two gradient-free optimization strategies considered. 
Similar to the previous task family, it can be observed that learning outcome improves with number of training generations.
Fig.~\ref{fig:learn_snn}(b) compares the readouts generated by a baseline reservoir and an ES-optimized reservoir to the target curves (SNN membrane potentials) in a single meta-testing task. As is evident from the query period, the optimized system is considerably better in reproducing the fluctuating dynamics of the membrane potential, which governs the spiking behaviors of the SNN readout neurons. 

To assess the learning gain from the L2L process, Table~\ref{table:snn} compares the NRMSE for Task 2 with and without meta-learning and shows substantial improvement with optimized reservoirs, up to $\simeq 50$\% in the case of ES optimization. 
Table~\ref{table:snn} and Fig.~\ref{fig:learn_snn}(a) also indicate that the learning outcome converges to a better result with ES compared to SA optimization.
This is because this task family (Task 2) employs more hyperparameters than the Volterra task family (Task 1) and ES typically performs better for a larger parameter space while SA is more suitable for stepping out of local minima.


\section{Discussion}
Previous studies have demonstrated the memory capacity of memristive reservoirs as well as their capability to generate dynamical features \cite{Sillin2013, Fu2020, Zhu2021information}. 
The results here suggest that memristive reservoirs are able to combine these properties together to enable learning of the rich, nonlinear time-delayed dynamics embedded by the Volterra filter. 
Furthermore, the L2L framework consistently determined that the optimal initial state for the memristive reservoir was at what can be described as the `edge of formation'. 
This point exists between where the memristive components first become activated (cf. Fig.~\ref{fig:snapshots}, $T_0$=2.0\,s), and before an exponential cascade of parallel paths form (cf. Fig.~\ref{fig:snapshots}, $T_0$=2.3\,s). 
The conductance exponentially ramps up at the time of initial formation, and saturates
as more parallel pathways form. The intermediate internal state, i.e., the `edge of formation' was meta-learnt as the optimal starting point prior to training in the outer loop.

This result opens up deeper insights to how memristive reservoirs can be optimally used in computation. 
It is somewhat intuitive that the reservoir does not perform at the time of initial formation, because the exponentially ramping conductance increase is highly unstable and challenging to control.
The opposite problem exists after conductive pathways are formed, where switching has a negligible impact on the dynamics of the nanowire network. 
The `edge of formation' can be thought of as a linear region in small-signal analysis, that fosters a controllable learning environment optimal for learning the rich dynamics of higher-order systems.

Additionally, this work also demonstrates that memristive reservoirs can learn the fluctuating dynamics of SNN membrane potentials.
This is possible because, as shown in a previous study \cite{Hochstetter2021}, memristive switching in a heterogeneous, recurrent network produces fluctuating dynamics that resemble action potentials which, when thresholded, generate spikes.
In other words, the internal dynamics of memristive reservoirs effectively encode spike-like features into continuous input signals.


An immediate implication of this result is that memristive reservoirs have the potential to serve as an interface between continuous data streams and spike-based computing paradigms, which could substantially improve the workflow for SNN applications.
The use of trainable spike-based embeddings has offered an alternative approach to classical rate and temporal encoders in many recent SNN works, but relies on gradient-based optimization to compress data into more efficient spike-based representations~\cite{zhu2023spikegpt, dold2022relational, zhang2023sscae}.
These results demonstrate how native internal dynamics, constrained by physics, naturally give rise to embeddings that can ultimately reconstruct signals in the context of gradient-free meta-learning.

\section{Conclusion}

This study shows that the L2L framework can effectively adjust the hyperparameters of memristive reservoirs, attaining a similar optimal regime as found by a manual search in previous studies.
Moreover, we demonstrated that the learning capability of the system can be extended and optimized for a family of related tasks, rather than being limited to a single task. 
This approach could pave the way for highly adaptive learning tasks based on real-world settings, similar to how  biological brains can learn quickly from limited examples.
Our finding that memristive reservoirs have the ability to reproduce SNN membrane potentials is significant because of the potential to use them in place of spike-based embeddings or encoders.
In particular, membrane potential dynamics in complex tasks may be encoded using memristive reservoirs as a more efficient mode of compression.

The memristive reservoirs studied here are motivated by physical reservoirs using self-assembled nanowire networks.
The neuromorphic properties of these networks are influenced by various factors such as nanowire density, diameter, average length and amount of dielectric. 
Similar to how evolutionary optimization schemes can be utilized to design and train neuromorphic systems \cite{Schuman2020}, it may be possible in the future to exploit the L2L framework to find the optimal nanowire networks for specific task families, effectively realizing a `learn-to-build' approach. 

\begin{acks}
The authors acknowledge the complimentary computing resources provided by Google Cloud.
The authors would like to thank Anand Subramoney for inspirational discussions on the L2L framework.
R.Z. is supported by the PREA scholarship from the University of Sydney.
\end{acks}

\bibliographystyle{ACM-Reference-Format}
\bibliography{references}

\end{document}